\documentclass[twocolumn,secnumarabic,amssymb, nobibnotes, aps, prd]{revtex4-1}

\usepackage{graphicx}

\begin{document}

\title{Characterization of Anomalous Forces in Dielectric Rotors}%

\author{Elio B. Porcelli$^1$}%
\email[Corresponding Author E-Mail: ]{elioporcelli@h4dscientific.com}
\author{Victo S. Filho$^1$}
\affiliation{$^1$H4D Scientific Research Laboratory, 1100 B15 Av. Sargento Geraldo, Santana, SP 04674-225 
}
\date{September 2020}%

\begin{abstract} 
We performed several measurements of anomalous forces on a dielectric rotor in operation, subjected to high voltage. 
The device operated under constant and intense angular velocity for each high voltage applied. The measurements were made 
in the similar way than an analogue magnetic gyroscope, by considering clockwise and counterclockwise rotations. We found 
that there are significant weight reduction on the device in the clockwise case, with one order of magnitude higher than 
the magnetic case. In addition, we detected a similar asymmetry in the observation of the effect, that is, there are 
smaller results for the anomalous forces in counterclockwise rotation. We also propose a theoretical model to explain 
the quantitative effect based on average values of macroscopic observables of the device rotation and concluded that 
it is consistent with the experimental results. 
\end{abstract}

\maketitle

\section{Introduction}
\label{sec:introduction}

It is well known from literature that gyroscopes can be applied in a lot of technological applications, as in 
systems of localization on Earth (GPS)~\cite{Bruckner,Gao,Najafi}, in electronic games~\cite{Allan,Ionescu}, in systems of 
navigation~\cite{Venka} and other cases, based on the physical properties related to the inertia of such devices. 
Besides, there is much interest in the research for new and more modern applications. In such a context, relevant reports 
have been made to the cases of gyroscopes in very small scales to be used like sensors~\cite{Soder}, in atomic 
scale~\cite{Fang}, in automotive cases to avoid vibrations~\cite{Acar}, in space navigation~\cite{Roser} and in 
other cases~\cite{Antonello}. 

The applications involving gyroscopes are in general related to guiding, orientation or tracking the systems containing 
the device. However, in a relatively recent work~\cite{Hay1} it was also investigated a new and novel property presented 
by Hayasaka's group: their reduction of weight when they are subjected to high angular velocity and clockwise rotation. 
Further it was also found that there was an asymmetry in this property, that is, if the gyroscopes rotated 
in the opposite direction (counterclockwise) that weight reduction completely vanished. After that discovery, 
the same research group implemented measurements in a new setup in which the system also rotated in 
high angular frequency but in free fall~\cite{Hay2}. It was found the same effect, that is, weight 
reduction for right rotation and null effect for the left one.   

The novel results motivated other research groups to reproduce the effect. However, it was not found evidences 
of anomalous forces, as in the experimental works reported in Refs.~\cite{Faller, Quinn, Nitschke}. The negative 
experimental results obtained from those groups discouraged further investigations and hence anyone proceeded in 
a more profound investigation of the phenomenon, which remained forgotten by a relatively long time. 

Despite of last reports, we started to investigate the phenomenon and reported our conclusions in our more 
recent work in the field~\cite{Elio2019}, showing that all the other papers published after Hayasaka's 
experimental work did not reproduce the same conditions reported in Ref.~\cite{Hay1}, so that this 
motivated us to theoretically explain it and implement some experimental measurements on such setups 
to check by ourselves if gyroscopes could really present both effects, that is, the asymmetry for the 
two directions of rotation and the weight reduction in the device. Such a motivation was based on some 
relevant main reasons: the earlier experimental works used rotors made of brass, a material of 
low magnetic permeability in relation to the one used in Ref.~\cite{Hay1} (silicon steel) and 
in the theoretical model proposed in Ref.~\cite{Elio2019} the role of macroscopic collective 
effect of the magnetic dipoles constituting the rotor is more than fundamental. In addition 
to the material, some alternative experiments were similar to the former but they differed 
from it in some other important features: the masses of the gyroscope rotors were 
substantially larger; the gyroscopes were spun in closed containers, but the devices 
were not operating in vacuum, the maximum rotational frequencies were lower and 
the rotors were not electrically driven but air driven. So, it is surprising 
that, with so many differences, one could conclude that there was an effective 
reproduction of the original experiments. In fact, the many factors indicated 
could surely avoid the clear manifestation of so weak forces in the operation 
of the devices, so that the results could not be really reproduced. 

The theoretical model presented in Ref.~\cite{Elio2019} could successfully explain 
the magnitude of such forces for the setup described in Ref.~\cite{Hay1}. In order 
to experimentally investigate the existence of such an effect in devices operating 
in rotation, we decided to establish a new experimental setup involving a rotor but 
considering a new physical configuration. As in our earlier works, we realized that 
magnetic dipoles and electric dipoles can both present such anomalies in the context 
of the generalized quantum entanglement (GQE), we implemented to perform some 
measurements of anomalous forces for a electrostatic material because in some 
of our previous works~\cite{Elio2015,Elio2016a,Elio2016b,Elio2020a,Elio2018d} 
we verified strong magnitudes of forces involving this kind of material 
subjected to high voltage.The effect detected in the gyroscopes cannot 
be successfully associated to any conventional theory. So, we elaborated 
the theoretical description of such forces in the experiments by means 
of a theoretical model based on GQE, by considering as quantum 
witnesses the electric permittivity and as macroscopic observable 
the angular momentum of the dielectric rotors. The basic conjecture 
relies on that macroscopic observables can present influence in their magnitudes 
by a huge collection of microscopic particles (as the magnetic dipoles in rotors 
of gyroscopes) has been checked in a lot of physical systems and presented 
consistency with experiments, as capacitors~\cite{Elio2015,Elio2016a,Elio2016b,Elio2020a}, 
magnetic cores~\cite{Elio2017a}, superconductors~\cite{Elio2017b,Elio2018a}, laser 
diodes~\cite{Elio2017c}, piezoelectric devices~\cite{Elio2018b,Elio2018c,Elio2020b} 
and electromagnetic drives~\cite{Elio2018d,Elio2019b}. 
Based on that idea, we conceived an analogue to the gyroscope described in Ref.~\cite{Hay1}, 
a rotor made of dielectric material, due to the higher magnitudes of electric field  
which could enhance the intensity of such anomalous forces and determine their existence 
doubtless. So, in this work we intend as main objectives to measure significant values of 
such forces as to demonstrate the actual existence of the phenomenon and propose a  
consistent theoretical explanation for such an effect. 

In next section, we describe our experimental setup and the measurements realized in our 
study. Next, we discuss in the section \ref{theory} a theoretical model that can describe 
the novel phenomena of induction of anomalous forces generated in the dielectric rotor 
under right rotations in rest frame of the laboratory. In the section \ref{results}, 
we present the experimental values of forces obtained and compare them with our 
theoretical calculations. At last, in section \ref{concl} we present our main 
conclusions and final remarks. 
\section{Experimental Work}
\label{exp}

In this section, we first describe our experimental setup in details, including the 
measuring devices used. Hence we explain all the procedures adopted to measure the 
existence and the magnitude of the anomalous forces, by means of high intense rotation 
of the device constituted by electric dipoles. That device analogous to the gyroscope 
reported in Ref.~\cite{Hay1} was formed by a disc with radial electric field coupled 
to the engine and to the voltage source, subjected to high voltage and high angular 
frequency. 

\subsection{Experimental Setup}

In order to perform measurements of force induced on the own device or induced at a 
distance by a dielectric disc in high rotation and subjected to high voltage, we 
establish an experimental setup relatively simple but highly efficient which 
allows us to rotate the disc in both directions by providing to the device 
an intense electric power via an electric power supply. Such an electric 
power source generates the variation of electric voltage without significant 
leakages of current during the operation of the rotor. This is a critical 
point because the connection of the device with the electric power supply 
or source is very difficult to implement accordingly. 

\begin{figure}[!t]
\centerline{\includegraphics[width=10cm]{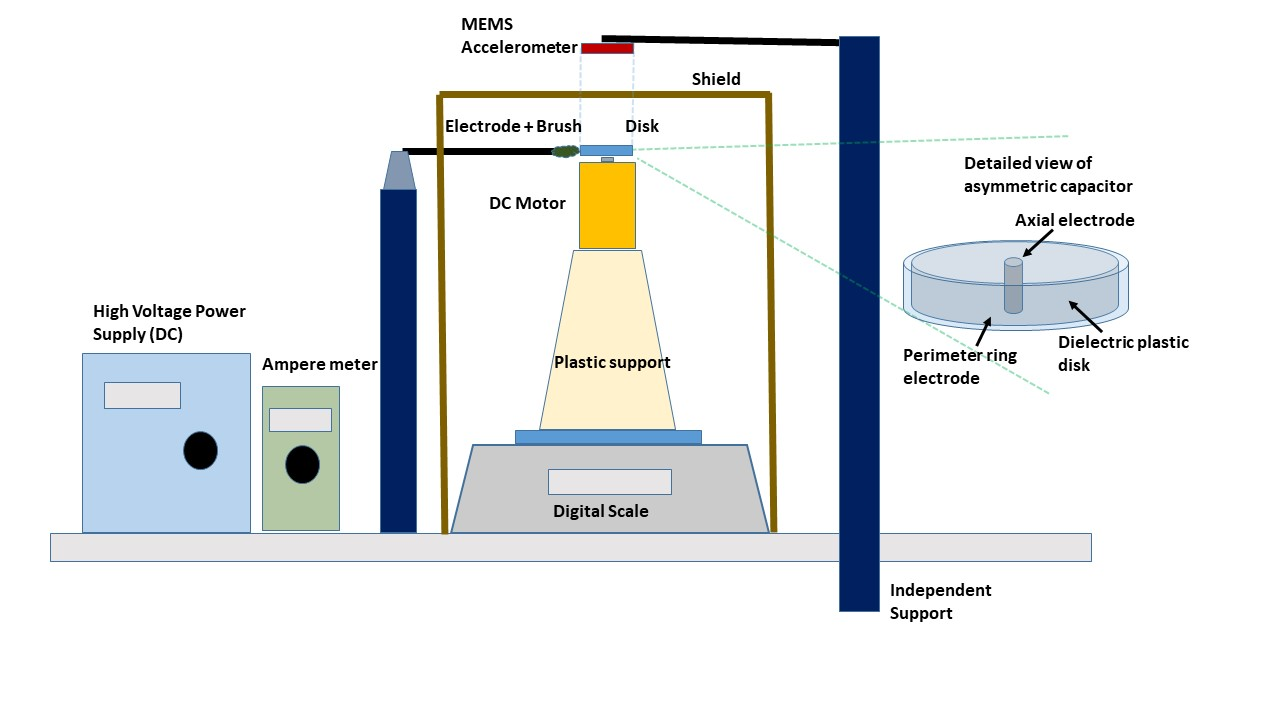}}
\caption{Physical diagram of the experimental setup used in our experimental 
work. In the center, we can see the dielectric rotor coupled to the motor, 
which in turn is supported on a cone placed on the tray of a scale. We also 
show the measuring equipments used in the experiment and other components 
of the assembly explained in the text. }
\label{f1}
\end{figure}

In Fig.~\ref{f1}, we show a physical diagram of the experimental setup used in our 
experimental work, including the measuring devices needed to 
detect the anomalous forces. 
As seen in that figure, the main part of the rotor is composed by a disc made of acrylic (see it in 
the detail at right), with metallic external border and a metallic ring in its central axis. The disc has 
diameter 0.04m, radius of its central ring 0.002~m, thickness 0.0025~m and mass approximately equal to 4~g. 
The perimeter of the disc is enclosed by an aluminum tape. The layer of aluminum measures 35$\mu$m and it 
is covered by an external insulated layer with 60$\mu$m. The internal metallic central ring is 
mechanically coupled to the metallic axis of a small DC engine (motor) responsible by the rotation 
of the entire set. The acrylic disc involved in the external border by a metallic ring and in the internal 
border by a metallic axis of the motor constitutes an asymmetric capacitor with symmetry axis placed in the 
vertical direction. It is rotated by the electric DC motor placed on a plastic cone, which in 
turn is placed on the scale tray, so that the disc was away from it, at a 22~cm distance, avoiding in 
this way any significant electrostatic interaction. 
The DC electric motor has a cylindrical plastic casing made by PVC film layers for electrical insulation and 
the scale tray that supports the motor-disk assembly has layers of EVA to damp the mechanical vibrations of 
the motor and for electrostatic insulation of the scale. 
The feeding wires of the motor and the electric connection of the dielectric disc axis are doubly insulated, 
placed apart and linked to supports to significantly decrease the electrostatic interaction among them. 

The Figure~\ref{f1} shows the assembly established in the experiments, including the wires, the 
battery, the switch key and the digital scale model BEL Mark 303, used to measure variations of 
weight of the order of 1~mgf of the disc-motor-support assembly, with weight approximately equal 
to 140~gf. 

The source of variable DC high tension was used to feed the dielectric disc by means of two brushes which are 
electrically connected to the axis and the external ring, each of them with different polarity. An accelerometer 
model Vectornav VN-100S Rugged was placed with its sensor Z vertically aligned with the symmetry axis 31~cm above 
it and supported by an independent support to avoid mechanical vibrations of the engine. The accelerometer was 
connected to the USB output of the computer by means of an insulated and armored wire. The computer executed a 
software of the accelerometer to register the reading of acceleration. Further a support was placed between 
the accelerometer and the disc to rule out to the maximum the possibility of wind, electrostatic interaction 
or acoustical noises. In other words, we take into account in the assembly all procedures to reduce or 
eliminate the electrostatic acoustic and mechanic interactions on the disc, the digital scale, the 
accelerometer, high tension source, feeding wires and external environment. 

\begin{figure}[!t]
\centerline{\includegraphics[width=10cm]{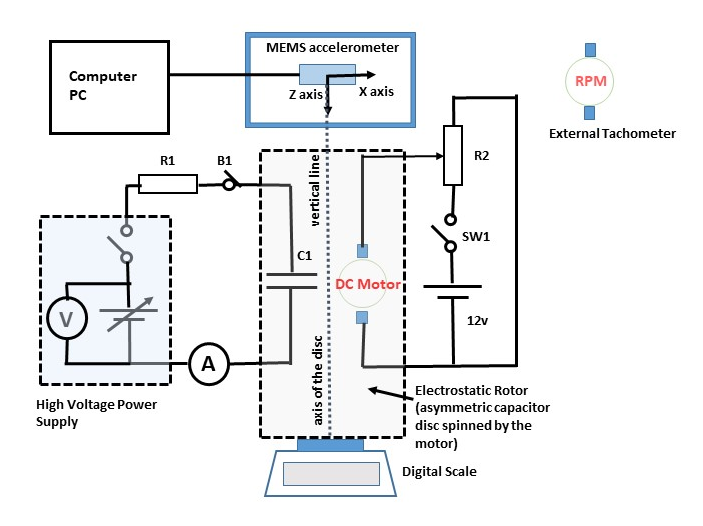}}
\caption{Scheme of the electric circuit of our experimental setup implemented to the measurements of anomalous 
forces on the dielectric rotor.}
\label{f2}
\end{figure}

The Figure~\ref{f2} shows a schematic draw of the electric circuit of the assembly or the electric diagram of 
the interconnection of the elements used in the experimental setup. Note that the DC electric motor of high 
rotation has its angular velocity adjusted by the rheostat R2 and its feeding provided by the 12~V battery, 
beside its activation to be made by the SW1 key. One can also observe that the disc (asymmetric capacitor C1) 
is fed by the high tension source which has a voltage adjust, an activation key and an internal volt meter through 
the brush B1 and the resistor R1 in the circuit. The latter corresponds to the insulating layer made of aluminum 
tape which constitutes the external ring of the disc. The charge current and leakage is controlled by the ampere meter 
linked in series with C1. 

In order to control the magnitude of the angular velocity of the dielectric disc was used an external digital tachometer 
Hikari model HDT-228 (with photo sensor and laser emitter), as detailed at the right top in the Figure~\ref{f2}. 

It is relevant to report that all the experimental measurements to detect auto-induction or anomalous forces on 
the own rotor were made by using the digital scale and all the measurements involving forces induced at a 
distance were made by means of the accelerometer. 

In the following, we describe our main experimental measurements made with such an experimental assembly. 

\subsection{Measurements of forces} 

We first made some measurements of weight variation or anomalous forces for different voltages applied on the device. 
During the execution of the experiments one realized that it was needed the application of higher values of 
tension to obtain weight variations overcoming all the noises or interferences, so that the effect was 
perceptible. Then, higher magnitudes of tension were applied, as shown in Table~\ref{tab1}. Table~\ref{tab1} shows 
three measurements of weight variation, by comparing with previous values of that during the 
application of high voltage in the dielectric disc in clockwise rotation at an angular 
velocity of 20000~rpm. We made three measurements to each level of voltage, from 
15300~V to 18300~V, which indicated that the weight loss of the disc is 
proportional to the magnitude of the applied voltage. 

For all the measurements shown in Table~\ref{tab1}, the experimental setup was the same in which the brush did the 
contact with a 5$^{\textnormal{o}}$ arc of the external ring of the dielectric disc whose external layer was 
insulated. Hence, only a small fraction (5/360) of the dielectric dipoles of the disc were effectively 
polarized. Therefore, we made an additional cover of the outer ring of the disc with aluminum foil. In 
this way, the loads were distributed throughout their external perimeter in the subsequent tests, thus 
optimizing the system so that all the dipoles inside the disk volume could be polarized.
\begin{table*}
\caption{Reduction of weight as function of the increasing of the voltage for clockwise rotation of the dielectric disc. 
The quantity $\Delta W$ in 2nd column represents the average weight loss measured. The abbreviation Diff. represents 
the difference between the maximum and minimum values obtained for each experimental test. LC indicates the 
Leakage Current in the experiments for any test implemented.}
\setlength{\tabcolsep}{3pt}
\begin{tabular}{|l|l|l|l|l|l|l|l|}
\hline
Voltage  &	$\Delta W$ & Test~1(g) Max,  &	LC($\mu$A)  & 
Test~2(g) Max,   &	LC($\mu$A)  & Test~3(g) Max,  & 	
LC($\mu$A) \\
(V) & (g) & Min \& Diff. & Max, Min & Min \& Diff. & Max, Min & Min \& Diff. & Max, Min \\
\hline
15300	&	   -     &  142.031	& 5.3	    & 142.062	& 4.7	    & 142.064 &	4.4 \\
      &	   -     &  142.023	& 3.0	    & 142.027	& 3.0	    & 141.985	& 3.0 \\
	-   &  0.041	 &   0.008  & -       &   0.035 &	-       &   0.079	& -	\\
\hline 
16300	&	   -     &  142.151	& 5.2	    & 142.067	& 5.5	    & 142.076 &	6.6 \\
      &	   -     &  142.060	& 3.7	    & 142.051	& 3.6	    & 142.021	& 3.5 \\
	-   &  0.054	 &   0.091  & -       &   0.016 &	-       &   0.055	& -	\\
\hline
17300	&	   -     &  142.200	& 7.1	    & 142.184	& 6.8	    & 142.166 &	6.3 \\
      &	   -     &  142.095	& 4.5	    & 142.108	& 4.2	    & 142.123	& 4.5 \\
	-   &  0.075	 &   0.105  & -       &   0.076 &	-       &   0.043	& -	\\
\hline       
18300	&	   -     &  142.351	& 8.4	    & 142.312	& 8.5	    & 142.291 &	7.9 \\
      &	   -     &  142.252	& 5.2	    & 142.177	& 5.5	    & 142.184	& 5.5 \\
	-   &  0.114	 &   0.099  & -       &   0.135 &	-       &   0.107	& -	\\
\hline
\end{tabular}
\label{tab1}
\end{table*}

This led to an increase in the magnitude of the measured weight loss, which reached 1.2~g, that is, 
more than 1/4 of the disc weight. In such measurement, the measured average current (116.9 $\mu$A) 
had a magnitude much greater than that of the currents obtained from the measurements indicated in 
Table~\ref{tab1} (from 3 $\mu$A to 8.5 $\mu$A). Table~\ref{tab5} shows the parameters for this measurement.

\subsection{Tests of Symmetry}
Experiments with a magnetic rotor indicated an asymmetry in the loss of weight of the device when it rotated in 
clockwise direction, but no significant loss in the opposite direction. The explanation for that asymmetry 
can be well understood in our previous work, in a qualitative way, in the framework of GQE~\cite{Elio2019}. 

In order to answer if the same novel effect occurs in a dielectric disc, we performed three batteries of 
experimental tests at clockwise rotation and also three of them at counterclockwise one. Each battery was 
made with six different values of DC voltage, from 10550~V to 15500~V, keeping the same angular velocity 
in all the configurations (around 19000~rpm to 22000~rpm). The experimental tests had the following 
procedure: in a first step, we measure the first variation of weight by the digital scale after the 
activation of the high voltage, that is, the source supply was turned on. Tables~\ref{tab2a} and 
\ref{tab2b} shows the values of weight immediately after its activation and the values of weight which 
stabilized after the activation in all of thirty six tests realized. It is notable that there was 
weight reduction for all the measurements in clockwise rotation of the dielectric disc, as one 
can realize in the experimental data given in Table~\ref{tab2a}; however, in the counterclockwise 
rotation there was both increasing and decreasing of weight, that is, the behavior was undefined 
and besides the weight variation presented one order of magnitude lower than those in opposite 
direction, as one can realize in the experimental data given in Table~\ref{tab2b}. In other 
words, the asymmetry was clear also in the case of dielectric rotor here conceived. 

\begin{table*}
\caption{Analysis of variation of the weight according to the voltage applied for the clockwise rotation.}
\setlength{\tabcolsep}{3pt}
\begin{tabular}{|p{35pt}|p{100pt}|p{59pt}|p{59pt}|p{59pt}|}
\hline
Voltage (V)	& Leakage Current~($\mu$A) \& Average $\Delta W$(g) & 
Test~1(g) Max Min Diff.	& Test~2(g) Max Min Diff.	& Test~3(g) Max Min Diff.\\
\hline
10500	& 0.4	& 126.322	& 126.278	& 126.354 \\
  & - &		126.201	& 126.206	& 126.236 \\
  &	-0.10367 & -0.121	& -0.072 & -0.118 \\
\hline
11500	&	0.6	&	126.339	&	126.201	&	126.174 \\
 &	- & 126.116	& 126.134	& 126.096 \\
 &	-0.12267 & -0.223 & -0.067 & -0.078 \\
\hline
12500	& 0.7	& 126.008	& 126.191	& 126.246 \\
 &	- & 125.067	& 126.093	& 126.188 \\
 & -0.36567 & -0.941	& -0.098	& -0.058 \\
\hline 
13500	& 0.7	& 126.389	& 126.465 &	126.435\\
 & - &		126.165	& 126.197	& 126.13\\
 &	-0.26567 & -0.224 & -0.268	& -0.305\\
\hline 
14500	& 1.1	& 126.553	& 126.56	& 126.534\\
 & - &		126.298 &	126.383	& 126.506\\
 &	-0.1533	& -0.255	& -0.177 &	-0.028\\
\hline
15500	& 1.4	& 126.699	& 126.623	& 126.644 \\
 &	- & 126.497	  & 126.567 	& 126.542 \\
 &	-0.12	& -0.202 & -0.056	& -0.102\\
\hline
\end{tabular}
\label{tab2a}
\end{table*}

\begin{table*}
\caption{Analysis of variation of the weight as function of the voltage applied for the counterclockwise rotation.}
\setlength{\tabcolsep}{3pt}
\begin{tabular}{|p{35pt}|p{100pt}|p{50pt}|p{50pt}|p{50pt}|}
\hline
Voltage (V)	& Leakage current ($\mu$A) \& Average $\Delta W$(g) & 
Test~1(g) Max Min Diff.	& Test~2(g) Max Min Diff.	& Test~3(g) Max Min Diff.\\
\hline
10500	& 0.4	& 117.007 &	117.288	& 117.521\\
  & - &	 117.28	& 117.575	& 117.62 \\
  &	0.21967 & 0.273 &	0.287	& 0.099 \\
\hline
11500	&	0.6	&	118.621 &	118.966	& 118.29	 \\
 &	- &  118.798	& 119.204	& 118.547 \\
 &	0.224 & 0.177 &	0.238	& 0.257	 \\
\hline
12500	& 0.9	&  119.99 &	120.065	& 120.045	 \\
 &	- & 119.965	& 120.021	& 120.195 \\
 & 0.027 & -0.025 &-0.044	& 0.15 \\
\hline 
13500	& 1.0	& 120.357 & 120.432 &	120.374 \\
 & - &	120.37 & 120.398	& 120.381 \\
 &	-0.00467 & 0.013	& -0.034	& 0.007	 \\
\hline 
14500	& 1.3	& 120.639 &	120.619	& 120.763 \\
 & - &	120.606	& 120.71	& 120.793 \\
 &	0.02933	& -0.033 &	0.091	& 0.03	 \\
\hline
15500	& 2.1	& 120.932	& 120.955	& 120.989 \\
 &	- & 120.896	& 120.909	& 120.905	 \\
 &	-0.05533	& -0.036	& -0.046 & -0.084\\
\hline
\end{tabular}
\label{tab2b}
\end{table*}

\subsection{Dependence on the Angular Velocity} 

Other sequence of measurements in our experimental work was also implemented to the case of different 
angular velocities. In order to verify if the dielectric disc could present a weight variation which 
changed as a function of the angular velocity, we performed some different measurements of the weight 
reduction as a function of the high tension when it is activated and deactivated to all the eight 
different levels of angular velocity, as shown in Table~\ref{tab3}. The angular velocity was 
adjusted as a function of the DC tension feeding the motor and it was measured with the external 
tachometer. There was a certain variation of angular velocity even for the same values of DC tension. 
The minimum and maximum values were also indicated in Table~\ref{tab3}. Analyzing the correlation 
between the weight reduction and the average angular velocity, we did not detect any correspondence 
or correlation between the weight reduction $\Delta W$ and the angular velocity for a high and 
constant value of tension applied of 12800~V. 

\begin{table*}
\caption{Analysis of variation of weight according to the variation of angular speed.}
\label{table}
\setlength{\tabcolsep}{3pt}
\begin{tabular}{|c|c|c|c|c|}
\hline
DC motor & Minimum rotation &	Maximum rotation & Average	& $\Delta W$ \\
voltage (V) & (rpm) & (rpm) & (rpm) & (g)\\
\hline
3.0	   & 7078	& 13266	& 10172	& -0.0495\\
\hline
4.8	 & 1334	& 22097	& 8947.5	& 0.24425\\
\hline
5.8	& 5294	& 21872	& 11479.75	& -0.15525\\
\hline
6.2	& 1296	& 13740	& 7518 & 	0.05525\\
\hline
6.6	& 271.6	& 36902	& 16144.2	& -0.0875\\
\hline
7.2	& 19035	& 19369	& 19202	& -0.338\\
\hline
9	& 14100	& 24323	& 19211.5	& -0.03925\\
\hline
10.8 & 	4234	& 37461	& 22994.75	& -0.04525\\
\hline
\end{tabular}
\label{tab3}
\end{table*}

\subsection{Experiments of Induction at a Distance}

Thirty measurements were also made to check if the dielectric disk 
could induce forces at a distance in such a way that gravitational 
acceleration measurements undergone changes detectable by an 
accelerometer positioned with its z sensor 31 cm above the 
axis of symmetry of the disk. Such changes in the gravitational 
acceleration cannot be explained by trivial local interactions 
such as electrostatic, acoustic or mechanical (wind) in view 
of the shield that was interposed between the disc and the 
accelerometer.

Table~\ref{tab4} shows a summary of the 30 measurements made, 6 measurements
for each of the five lines shown, each showing the direction of rotation 
of the disk and if the body of the accelerometer was aligned 
within the vertical projection of the circular area of ​​the disk 
or was misaligned or outside the same projection.

Each of the 30 measurements corresponds to 80 samples obtained by 
the accelerometer in a period of 2~s immediately prior to the 
high voltage activation period and 2~s immediately after the 
same activation.

Table~\ref{tab4} shows that the dielectric disc induced a force whose vector 
points downwards on the activation of high voltage (12800~V) and 
clockwise rotation of 20000~rpm in such a way that the gravitational 
acceleration measured by the accelerometer z-sensor has decreased. 
Such a decrease was more than four times smaller than when the 
accelerometer body was misaligned with respect to the dielectric 
disk, which denotes that the distance induction has a collimation 
feature or a dependence on the geometry.
\begin{table*}
\caption{Analysis of the variation of acceleration induced by the disc at a distance. In first 
column, CW means clockwise and CCW means counterclockwise. In third and fifth columns, D$_a$ represents 
the standard deviation of the acceleration. }
\setlength{\tabcolsep}{3pt}
\begin{tabular}{|p{100pt}|p{35pt}|p{35pt}|p{35pt}|p{35pt}|p{50pt}|} 
\hline
 - & Voltage "off" & Voltage \ \ "off" & Voltage \ \ "on" & Voltage \ \ "on" & Difference \\
\hline
Direction of rotation  	&	Average  &	D$_a$ & Average  & D$_a$ & Average  \\
\& Alignment & a(m/s$^2$) & (m/s$^2$) & a(m/s$^2$) & (m/s$^2$) & a(m/s$^2$) \\
\hline
CW - Aligned 			      & -9.1604 & 0.0623 & -9.1417	& 0.0563 & 0.0187 \\
\hline
CCW - Aligned	    & -9.1454 & 0.0423 & -9.1545	& 0.0393 & -0.0091 \\
\hline
CW - Misaligned 		    &	-9.1407	& 0.046	 & -9.1362	& 0.054	 & 0.0045 \\
\hline
CCW - Misaligned	&	-9.1427	& 0.039	 & -9.1445	& 0.0537 & -0.0018 \\
\hline
No rotation - Aligned			    & -9.1366	& 0.048	 & -9.1307	& 0.0363 & 0.0059 \\
\hline
\end{tabular}
\label{tab4}
\end{table*}

Both effects, that is, remote induction and self-induction of forces 
(where the forces are measured as a variation in the weight of the dielectric 
disk itself) in the experiments, are not well explained considering those  
trivial interactions such as electrostatic, acoustic, or mechanical, so 
they can be ruled out according to the procedures, protections and 
insulation that have been carried out. Such effects concerning 
to the generation of anomalous forces or weight reduction have been 
observed in other different physical systems, as in the case of weight reduction 
in capacitors~\cite{Musha} and in case of Podkletnov experiments involving 
superconductors~\cite{Podkletnov,Solomon,McCulloch} and some authors have 
proposed some possible explanations for that~\cite{Solomon,Zhu}. 
For instance, in Ref.~\cite{Zhu} it has been shown the conditions for the existence 
for the interaction between the electromagnetic and gravitational fields, so that 
the variation of gravitational acceleration could be measured.

So, we propose a possible theoretical explanation to the phenomenon. 
Next, we propose a theoretical model that can explain both qualitative and 
quantitative experimental results.

\section{Theoretical Model} 
\label{theory}
The theoretical model is based on the idea that the myriad of internal 
electric dipoles in the acrylic dielectric disc which undergoes a polarization 
of intense radial electric field between the external ring and the axis really 
affects the physical quantities of the system in macroscopic scale. In 
this condition, if the dielectric disc is under high rotation then the electric 
dipoles also undergo a precession effect~\cite{Talman}, so that either they 
oscillate around a vector pointing downwards if the rotation of the disc is 
clockwise (that is, the angular momentum of the disc points downwards 
in the vertical direction) or oscillates around a vector pointing 
upwards if the rotation of the disc is counterclockwise (or if  
the angular momentum of the disc aligned to the vertical direction 
points upwards). 

The model also considers the existence of generalized quantum 
entanglement (GQE) as a conjecture~\cite{Elio2017a} between the myriad 
of electric dipoles collectively synchronized in precession and the 
environment around the disc. In general, that non-local interaction 
is very weak, but in some special conditions such an interaction 
can become more intense and be perceptible. 

When the electric dipoles precess downwards, due to the non-local 
interaction with the planet below them, they undergo a 
reaction force upwards so that there is a decreasing in the 
weight of the disc. That same process was verified in the case 
of magnetic dipoles in magnetic rotors, according our previous 
work~\cite{Elio2019}. When the electric dipoles precess upwards, 
due to the non-local interaction with the atmosphere and the 
outer space above them, they undergo a reaction force downwards, 
but it is very weak because the density of the matter above the 
device is very small when compared with the density of the Earth. 
So, it is clear that there exists an asymmetry in the framework 
of GQE as an explanation to the anomalous forces arising in the 
device. In other words, the disc looses weight in clockwise 
rotation, but it does not effectively gain weight in 
counterclockwise rotation, exactly as occurs in the 
case of the gyroscope~\cite{Elio2019}. Such an asymmetry 
was in fact detected in our experiments whose data are 
presented in Tables~\ref{tab2a} and~\ref{tab2b}. 

Based on the model here proposed, it is possible to calculate 
the magnitude of the reaction force that the electric dipoles 
in precession can induce in the disc in clockwise rotation 
when subjected to a static radial electric field. In reality, 
each electric dipole in precession transfers a non-local force 
in the direction of the precession, which is the direction where 
points out the vector angular momentum of the disc. Hence, in 
first place, we must calculate the torque $\tau_i$ of the individual  
electric dipole $i$, whose magnitude of dipole momentum $p_i$ is 
about $p_i = $ 1.6 x 10$^{-29}$ Cm, according to Ref.~\cite{Housecroft,Podesta,Tro}. 
That magnitude is consistent with the electronic polarization of 
the dielectric, which is dominant in the device due to the lower 
value of the dielectric constant of the disc: $\epsilon_r = $ 3.22 
and the application of the static radial electric field which is 
not variable in time. 

The torque $\tau_i$ is easily calculated by 
\begin{equation}
\tau_i = p_i E \sin \theta, 
\end{equation}
in which $E$ is the magnitude of the electric field vector 
$\vec{E}$ which polarizes all the electric dipoles constituting 
the material of the dielectric. The electric field can be  
calculated according some parameters as the radial distance 
between the bigger electrode (ring involving the perimeter 
of the dielectric disc) and the smaller one (metallic axis 
in the center of the dielectric disc), the difference of 
electric potential $V$ between the electrodes (voltage 
applied) and the factor of field concentration $G$, 
according to Ref.~\cite{Fazi}, if we take into account 
that the system represents an asymmetric capacitor 
with electrodes of different dimensions. 
So, we can write the magnitude of the 
electric field as: 
\begin{equation}
E = \frac{V}{R_{ring}-R_{axis}} G(V), 
\label{eq1}
\end{equation}
in which $R_{ring}-R_{axis}$ is the difference between 
the values of radius of the external ring and the axis 
and $G$ is the gradient of the electric field, which 
can be calculated by 
\begin{equation}
G = \frac{R_{ring}}{R_{axis}}, 
\label{eq2}
\end{equation}
so that Eq.~(\ref{eq2}) represents the ratio involving 
the radii of the electrodes. 

In this way, it is possible to calculate the magnitude of 
the torque $\tau_i$ undergone by each dielectric dipole, by 
considering $\theta = \pi/2$, that is, the direction of the 
axes where the electric dipoles precess around is perpendicular 
to the direction of the radial electric field $\vec{E}$ in 
the bulk of the disc. 
As the magnitude of the torque is the contribution of all 
individual torques $\tau_i$ of the electric dipoles $i$, 
we can write as average torque $\tau$ = $n\tau_i$, in which 
$n$ is the number of dipoles. The next step is to calculate 
the quantity of electric dipoles $n$ of the dielectric disc. 
For that, it is needed to calculate the polarization $P$ of the 
dielectric~\cite{Woan,Mossotti,Clausius}, according to: 
\begin{equation}
\vec{P} = Z \epsilon_0 \frac{\epsilon_r-1}{\epsilon_r+2}\vec{E}V_0, 
\end{equation}
in which $\epsilon_r$ is the dielectric constant of the material 
(acrylic)~\cite{PPD}, $\epsilon_0$ is the dielectric constant of the vaccum, 
$\vec{E}$ is the effective electric field and $Z$ is a 
proportionality constant~\cite{Indulgar} which has value 3 to 
isotropic dielectric materials, but we here consider it in our 
model as $Z$ = 1. This value is considered different than 3 
due to the polarizability anisotropy of the polymer PMMA 
(acrylic) used in the dielectric disk~\cite{Pol1,Pol2}.  

The number of electric dipoles $n$ is calculated by~\cite{Jan}
\begin{equation}
n = \frac{|\vec{P}|}{|\vec{P_i}|}, 
\end{equation}
with 
\begin{equation}
\vec{P} = \sum_i \vec{P_i}, 
\end{equation}

Hence, it is possible to obtain the average total torque as 
$\overline{\tau} = n \tau_i$ and we can also calculate the 
value of the average force by
\begin{equation}
\overline{F} = \frac{\overline{\tau}}{r}, 
\end{equation}
in which $r$ is the radius of the dielectric disc. Such a force 
of reaction raises due to the precession of the dielectric 
dipoles originated from the electric field applied and the 
angular momentum of the disc. Such an effect would not be 
possible if it was not intermediated with the exterior 
(environment) by considering the preexistent generalized 
quantum entanglements. 

\section{Results} 
\label{results}
Our calculations indicate that there is a good agreement with 
the experimental data obtained from literature for most rotation 
frequencies measured. The theoretical model presented 
in section \ref{theory} is consistent with the experimental 
results given in Table~\ref{tab5} of section \ref{exp}. 

Table~\ref{tab5} shows the theoretical predictions of the 
magnitude reaction force according to the experimental parameters 
used in the calculations demonstrated up to this point and also 
the experimental measurements of that force, that is, the 
decreasing of the disc weight in clockwise rotation also 
indicated in Table~\ref{tab1}. 

{\small{
\begin{table*}
\caption{Theoretical forecast according to the parameters of the experiments.}
\setlength{\tabcolsep}{3pt}
\begin{tabular}{|p{80pt}|l|l|l|l|l|}
\hline
Parameter	                       & 1st Runs	          & 2nd Runs         & 	3th Runs      	  & 4th Runs	               & 5th Runs \\
\hline
Energized portion of the ring    & 0.01389	          & 0.01389	         & 0.01389	          & 0.01389	                 & 1.0 \\
\hline
Electric dipole(Cm)	         & 1.6x10$^{-29}$     & 1.6x10$^{-29}$   & 1.6x10$^{-29}$     & 1.6x10$^{-29}$           & 1.6x10$^{-29}$\\
\hline
Effective voltage (V)            &	15300	            & 16300	           & 17300              & 18300	                   & 8378.88\\
\hline
$\epsilon_r$	                   & 3.22	              & 3.22	           & 3.22	              & 3.22	                   & 3.22\\
\hline 
$\epsilon_0$~(F/m)	             & 8.85x10$^{-12}$	  & 8.85x10$^{-12}$	 & 8.85x10$^{-12}$	  & 8.85x10$^{-12}$          & 8.85x10$^{-12}$\\
\hline
E field~(MV/m)	                 & 8.5	              & 9.056 	         & 9.61	              & 10.167 	                 & 4.655 \\
\hline
Polarization(Cm)	             &  1.38x10$^{-12}$  	&  1.47x10$^{-12}$  &   1.56x10$^{-12}$	  &   1.65x10$^{-12}$          &   5.45x10$^{-11}$\\
\hline
Big electrode(m)	               & 0.02	              & 0.02	           & 0.02	              & 0.02	                   & 0.02\\
\hline
Small electrode (m)	             & 0.002	            & 0.002	           & 0.002	            & 0.002	                   & 0.002\\
\hline
Thickness of \ \ \ \ \ \ dielectric disk(m) & 0.0025	            & 0.0025	         & 0.0025     	      & 0.0025	                 & 0.0025\\
\hline
Dielectric vol.(m$^3$)	      & 3.110x10$^{-6}$	  & 3.110x10$^{-6}$ & 3.110x10$^{-6}$	  & 3.110x10$^{-6}$	       & 3.110x10$^{-6}$\\
\hline
Gradient of field	          & 10	                & 10	             & 10	                & 10	                     & 10\\
\hline
Dipoles number	              &  8.638x10$^{16}$  	&  9.202x10$^{16}$ &  9.767x10$^{16}$	  &  1.033x10$^{17}$	       &  3.41x10$^{18}$\\
\hline
Torque of each dipole~(Nm)	    &  1.36x10$^{-22}$	    &  1.45x10$^{-22}$ &  1.54x10$^{-22}$  	&  1.63x10$^{-22}$	       &  7.45x10$^{-23}$\\
\hline
Torque~(Nm)	                    &  1.18x10$^{-5}$	  &  1.33x10$^{-5}$ &  1.50x10$^{-5}$ 	&  1.68x10$^{-5}$	       &  2.54x10$^{-4}$\\
\hline
Theoretical Force~(gf)	        & 0.0599              &	0.067998 	       & 0.0766 	          & 0.0857	                 & 1.2937\\
\hline
Average exp. force~(gf) & 0.041	      & 0.054	      & 0.075       &	0.114     	& 1.2\\
\hline
Minimum exp. force~(gf) & 0.008	      & 0.016       & 0.043      	& 0.099	      &  -  \\
\hline
Maximum exp. force~(gf) & 0.079      	& 0.091	      & 0.105	      & 0.135       &   -	\\
\hline
\end{tabular}
\label{tab5}
\end{table*}
}}

The parameters of the data sets from 1 to 4 are equal, with 
exception of the voltages applied (from 15300 V to 18300 V) 
by the contact brush (electrode) in 5$^{\textnormal{o}}$ of the 
arc of the ring, so that only a part of it (5/360) was 
energized. So, only a fraction of the volume of the dielectric 
disc had its internal electric dipoles polarized. The theoretical 
values of the magnitude of the force are consistent with the 
experimental results shown in all those data sets (1 to 4) if 
we consider that they are within the interval between the 
maximum and minimum measurements and they are also very close 
to the average values which were measured. 

In the data set 5, the parameters were equal to the earlier, but 
with an important exception, that is, all the perimeter of the 
ring was energized so that all the dielectric dipoles of the 
entire volume of the dielectric disc have been polarized. 

The result of that experimental procedure was a very high value 
of the magnitude of the force calculated in our model: 
$\overline{F}$~=~1.29367 gf. That value is very close to the 
value measured in our experiments: $F_{exp} \cong$  1.2 gf. Such 
a loss of weight represents about 1/4 of the total weight of the 
dielectric disc ($W \cong $~4~gf). 

It is important to emphasize that in all the experimental rounds 
shown in Table~\ref{tab5} there was loss of weight consistent 
with the same magnitude of angular velocity ($\omega$ = 20000 rpm) 
and the direction of disc rotation (clockwise). 

In summary, 
the Tables~\ref{tab2a} and \ref{tab2b} shows the weight variation 
with relation to the direction of rotation (right or left) 
to different voltages applied to the system, that is, 
from 10500 V to 15500 V. Such results allow us to conclude the 
existence of an actual asymmetry in the effect, according to 
the theory here described. Besides, for all voltages applied 
to the device there was weight reduction when the rotation 
was clockwise, in opposition to the case of counterclockwise 
rotation, in which no reduction was in fact detected in the 
accuracy considered (at least one order of magnitude lower 
than the positive case). 

The measurements performed in the experiments whose values can be 
seen in Tables~\ref{tab2a} and \ref{tab1} were more conclusive than 
the ones in Table~\ref{tab2b} because the voltages applied were higher 
so that the residual variations on the background as noises and 
mechanical variations of the rotor were overcome in the former case. 
Despite of that, it is doubtless the asymmetry observed in 
Tables~\ref{tab2a} and \ref{tab2b}. 

Table~\ref{tab3} shows measurements of force performed in clockwise 
rotation to different values of angular velocity by means of the 
change of electric tension feeding the engine which imposes the 
rotation of the electric disc. It is worth to report that the voltage 
applied on the ring was the same in all the measurements, e.g., 
11000~V. 

In most part of the measurements, there was a loss of weight in the 
right rotation, but there was not observation of increase in the 
variation of weight according to the increase of the angular 
velocity. In the model proposed in this paper and according to 
the theory of gyroscopes~\cite{Jan}, the precession velocity 
$\Omega$ of the electric dipoles depends on the torque $\tau$  
on them applied, when subjected to a radial electric field 
$\vec{E}$, and also depends on the angular momentum $L_i$ which are subjected to, 
due to the angular velocity of the dielectric disc, so that the 
angle of precession $\theta$ can vary in time according to: 
\begin{equation}
\Omega = \frac{\tau_i}{L_i \sin \theta} = \frac{d\theta}{dt}.
\end{equation}

Note that the torque which each electric dipole undergoes with the 
application of the electric field $\vec{E}$ is responsible for the 
magnitude of the force transferred to the exterior via non-local 
interactions and the direction of the force varies as a function 
of the precession angle and the latter depends on the angular 
momentum of each electric dipole due to the rotation of the 
dielectric disc. Therefore we conclude that the reaction force 
(or the weight reduction) does not depend on the angular velocity, 
confirming the experimental results shown in Table~\ref{tab3}. 
Such a new result does not occur in the case of magnetic 
rotors, as seen in our previous work~\cite{Elio2019} because 
the radial magnetic field increased in magnitude when the 
angular velocity of the rotor also increased. In the present 
case, the radial electric field remained unchanged in 
magnitude, independently the variation of the angular 
velocity. 

The theoretical model proposed in this paper indicates that 
the electric dipoles interact with the exterior environment 
(or, in other words, with their particles) due to the 
existence of the quantum entanglement property. When 
the particles undergo the torque of the electric field, 
they transfer forces to the external environment which 
depend on the direction of precession. Such a direction 
is either downwards when the disc rotation occurs in the 
clockwise direction or upwards when the rotation occurs in the 
counterclockwise direction. Such a phenomenon was experimentally 
verified, as one can be seen in Table~\ref{tab4}. 

We detected a collimated field of forces by means of an 
accelerometer placed above the dielectric disc by tens 
of centimeters. The accelerometer z-sensor  was also aligned with 
the symmetry axis of the disc and then we detected the 
decreasing of the gravity acceleration (or, in other 
words, its body undergoes a force downwards) when the 
disc rotated at right. Analogously, we detected an 
increase in the gravity acceleration (the body of the 
accelerometer undergoes a force upwards) when the disc 
rotated in the opposite direction. Such a behavior occurred 
even with the lack of alignment of the z-sensor in 
relation to the symmetry axis of the disc. In fact, there 
was a dislocation of the accelerometer in the Y-axis 
as to stay out the projection of the circular area of 
the disc in the z-axis. However, the variation of 
the gravitational acceleration measured was much smaller, so 
that we really verified that the force induced by 
the disc in rotation and under high tension presented 
as a feature the dependence on the collimation. Such 
a property or geometry of forces induced depends on 
the precession of the electric dipoles and deserves 
a more profound investigation in future works. \\

\section{Conclusion}
\label{concl}

In this paper we performed some experiments of force measurements in a device analogue to the gyroscope or 
magnetic rotor. We investigate three main novel features of our dielectric rotor: the existence of anomalous 
forces or weight reduction of the device, the asymmetry in the magnitude of those forces (significant higher forces 
only appeared for right rotations) and the possibility of induction of forces at a distance. We conclude 
that the three features do occur in our experimental dielectric rotor for high values of voltage applied to that   
device and high angular velocities. No dependence on the angular velocity was observed, an unique 
difference in relation to the analogue magnetic rotor previously reported in Ref.~\cite{Hay1}. 
We also propose an empirical theoretical model to explain the effect which was based on 
the conjecture of generalized quantum entanglement which provides us an expression 
indicating that the collective precession of the electric dipoles constituting 
the dielectric rotor can accurately describe the experimental data in all 
the cases. Hence we conclude that our theoretical model is consistent 
with the experimental results. It is notable to realize that in the 
best result we obtained approximately a 1.2~gf weight reduction of 
the device, corresponding to impressive 25\% of the device 
weight. 



\end{document}